# Current Web Application Development and Measurement Practices for Small Software Firms

Faudziah Ahmad[1], Fauziah Baharom[2] and Moath Husni[3]

[1,2,3] School of Computing, UUM College of Arts and Sciences, Universiti Utara Malaysia, Malaysia

[1]*fudz@uum.edu.my*, [2]*fauziah@uum.edu.my*, [3]*tarawneh80@yahoo.com*

**Abstract**
This paper discusses issues on current development and measurement practices that were identified from a pilot study conducted on Jordanian small software firms. The study was to investigate whether developers follow development and measurement best practices in web applications development. The analysis was conducted in two stages: first, grouping the development and measurement practices using variable clustering, and second, identifying the acceptance degree. Mean interval was used to determine the degree of acceptance. Hierarchal clustering was used to group the development and measurement practices. The actual findings of this survey will be used for building a new methodology for developing web applications in small software firms.

## 1. Introduction

Web application is defined as a "Web system which consists of Web server, network, HTTP and browser, in which user input (navigation and data input) affects the state of the business" [1]. In general, Web-based applications differ from other traditional applications in terms of high reliability, high usability, security, better technologies, shorter time to market, shorter product life cycles and continuous maintenance [2].

Eighty five percent of software companies that are involved with developing Web applications consist of small software firms [3]. Small software firms refers to any organization or company that has approximately 10 to 50 employees [4][5][6]. One of the problems that is faced by these companies is that they do not know or apply a standard or best practice when developing a web application [7][8][9].

A best practice is defined as a management or technical practices that has consistently demonstrated and should be taken in to account to improve one or more of productivity, cost, schedule, quality, user satisfaction and predictability of cost and schedule [10]. On the other hand, software measurement involves understanding, controlling, predicting and improving software development project which is useful for reducing defects, reducing rework and reducing cycle time. In order to be effective, the measurement process must be integrated to the whole process and not just applying on a specific stage in the development process [7][9].

Many researchers have highlighted the importance of following a standard or best practice on web development. However, several studies have shown that there has been a lack of awareness of deploying the important development practices during the development process. However, to date the actual web development and measurement practices have not been investigated [11].

In Jordan, many software firms are considered as small firms. Jordanian government has little knowledge on the quality of services or products provided by these small firms [8]. Therefore, an empirical study was conducted in Jordan to investigate current web application development and measurement practices in small software firms. This paper presents findings of the pilot study.

## 2. Methodology

The research was conducted in three stages: data coding, variable clustering and acceptance degree identification. Fig 1 shows the flow of the research.

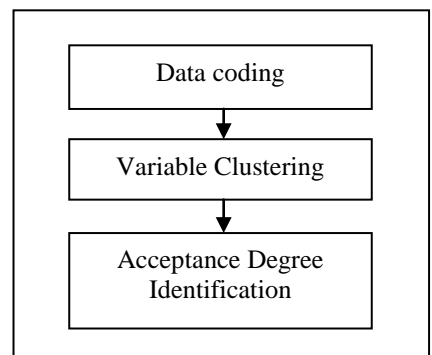

Fig. 1 Methodology

## 2.1 Data coding

During this stage, three activities were conducted: questionnaire construction, data collection, and data entry. Questionnaire was constructed based on literatures of web applications development and software development practices [8][15][16]. The questionnaire consisted of three sections: demographic information, development and measurement issues, and web application development and measurement practices. Besides, questionnaire, interviews were used to gather data. This paper discusses findings on the web application development and measurement practices. The first two parts have been discussed in [24].

In this pilot study, twenty three small software firms were selected randomly. Respondents were developers and managers of small software firms. Seventeen questionnaires were given to developers and six were given to managers. Each respondent answered the questionnaire with the researcher's guidance. Each question describes one practice. Table 1 shows examples of the practices in the questionnaire.

Data gathered from the questionnaires were then coded in SPSS version 14.0 (Statistical Package for Social Science) for analysis.

Table 1: Practices

| No | Practices | Variable |
|---|---|---|
| 1 | Development process was able to cope with time pressure | D1 |
| 2 | Development process was clarified, explained, and delegated the roles and responsibilities to team members | D2 |
| 3 | Development process was performed with minimum design and produce a prototype in a short time | D3 |
| … | … | … |
| 17 | There exist a procedure for maintaining awareness of the state-of-the-art in case of web engineering technology | D17 |

## 2.2 Variable clustering

Cluster analysis is a technique used for combining variables into groups. These groups are homogeneous i.e., variable inside the group are similar to each other. Variables in each group should be different from the other groups [22][23]. One of the commonly technique used for grouping variables that exist in SPSS is hierarchal clustering.

Clustering was conducted because of the difficulty to deal with large number of practices. Besides, clustering, factor analysis can also be used to group variables. However, this technique is not suitable for small sized data [20][21]. Cluster analysis was chosen for this purpose because it could deal with small data size. In this study, variables were clustered using hierarchal clustering and Wards method was used to determine the distance between each group and to determine which variable belongs to which group. Dendrogram was used to present clustering results.

## 2.3 Acceptance degree identification

In this stage, the acceptance degree was calculated using mean interval for each practice. It is meant to determine the extent of small software firms in Jordan apply important development and measurement practices during the development process. Mean was used because it takes into account all the values in the distribution, making it sensitive to extreme values [18]. Five Likert scales ranging from strongly disagree (value 1) to strongly agree (value 5) were used to describe the degree of acceptance for applying these practices in the development process of each company.

Results were calculated by getting the mean score and selecting the appropriate interval that represent the actual mean. An appropriate interval scale was required to represent all levels of acceptance.

The interval was calculated by the following equation ( Eq (1)):

$$\text{Appropriate interval} = (\text{number of scales} - 1) / \text{number of scales} \quad \text{------------------------} \quad Eq(1)$$

An example is
$$\text{Appropriate interval for the study} = (4/5) = 0.8$$

Scales representation for the degree of acceptance for each practice is shown in Table 2.

Table 2: Internal representations for the degree of acceptance

| Mean interval | Degree of acceptance |
|---|---|
| From 1 to 1.80 | Strongly Disagree |
| From 1.81 to 2.60 | Disagree |
| From 2.61 to 3.40 | Neutral( Don't Know) |
| From 3.41 to 4.20 | Agree |
| From 4.21 to 5 | Strongly Agree |

## 3. Findings

This section presents results on (i) variable clustering and (ii) acceptance degree.

### 3.1 Variable clustering

Fig 2 shows the results. The practices are grouped into seven clusters. Cluster1 consists of practices D6, D12 and D13. These practices are related to requirements. Cluster2 relates to quality issues and it consists of practices (D10 and D11). Cluster3 (D5, D14 and D15) relates to measurement practice. (D3, D7 and D17) are members of cluster4 and are related to web design. Cluster5 consists of (D4 and D16). These practices are related to management. Cluster 6 (D1 and D2) are related to the development process. Finally, D8 and D9 are grouped as Cluster 7 and is related to testing process.

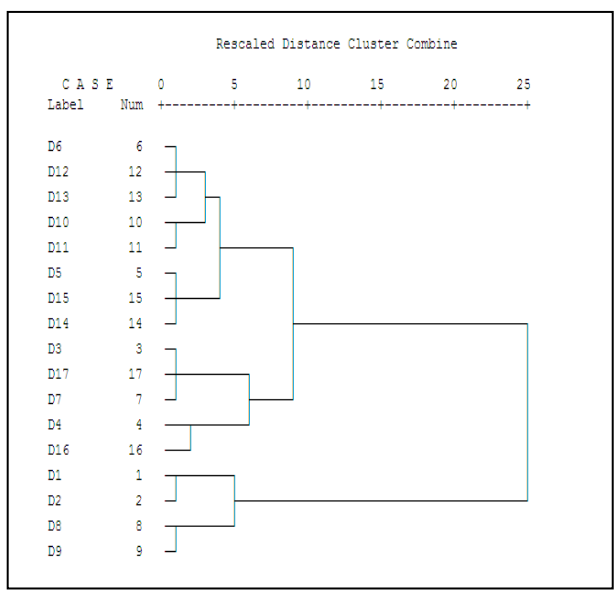

Fig. 2 Dendrogram.

Each practice in the cluster has a mean value that determines the degree of performing this practice during the development process. The acceptance degree of performing each practice was calculated by taking the mean value and matching it with the interval value representation in Table 2. Tables 3 to Table 9 show the mean value and degree of acceptance for all clusters (Cluster1 to 7).

Table 3: Requirement Practices

| Requirements Practices | Mean Value | Degree of Acceptance |
|---|---|---|
| User or and the manager are the direct sources for requirements (D6) | 1.96 | Disagree |
| Documented procedures are used for controlling requirements changing (D12) | 2.26 | Disagree |
| Change control function is established for each web project (D13) | 2.30 | Disagree |

Cluster1 Requirement Practices: This cluster is very important because it illustrates the way that organizations collect requirements. Table 3 illustrates that those three important practices has mean interval values between 1.81 and 2.60, indicating that developers of small software firms in Jordan did not follow the requirements of best practices.

Table 4: Quality Practices

| Quality Practices | Mean Value | Degree of Acceptance |
|---|---|---|
| Quality management standards are taken into consideration when developing web applications (D10) | 2.13 | Disagree |
| There are guidance on Software Quality Assurance while conducting the testing process (D11) | 2.48 | Disagree |

Cluster2 Quality Practices: this group of practices is related to the quality management (e.g. usability and user interface design) and quality assurance. Quality management is related with activities and tasks needed to maintain a desired level of excellence and quality assurance are related to how developers conduct testing. The aims are to identify whether they conduct the testing process themselves or by the users (under the guidance of software quality assurance team) and to investigate whether the developers pay attention to the quality management and standards. Results (Table 4) indicate that most developers are not concerned with quality management and assurance with the development process.

Table 5: Measurement Practices

| Measurement Practices | Mean Value | Degree of Acceptance |
|---|---|---|
| Project plan performed within the budget estimation (D5) | 2.35 | Disagree |
| Web applications size measures (such as "Lines of Source Code") are used (D14) | 2.04 | Disagree |
| Developers used measurement elements or standards to produce web development effort, schedule, and cost estimates (D15) | 1.87 | Disagree |

Cluster3 Measurement Practices: this type of practices is related to the method of product measuring in terms of budget, size, effort and schedule. These practices are aimed to reduce development cycle and minimize defects and rework. All practices (Table 5) in this study have the same degree of acceptance (disagree) during the development process despite knowing the importance of applying such practices within the development process. This means that they do not use any measurement during the development process – no assurance to a quality web application.

Table 6: Design Practices

| Design Practices | Mean Value | Degree of Acceptance |
|---|---|---|
| Development process performed with minimum design (D3) | 2.43 | Disagree |
| Design notations used in web design (D7) | 2.65 | Neutral |
| Procedure for maintaining design ensured the state-of-the-art of the web applications (D17) | 2.00 | Disagree |

Cluster4 Design Practices: design phase is a very important phase of the development cycle. A set of design practices must be taken into consideration to ensure that the development can be done in quickly and usable. As shown in Table 6 two practices "disagree" and one shows "neutral" acceptance. This means that the developers currently follow a complex and time consuming way of design for developing a web application.

Table 7: Management Practices

| Management Practices | Mean Value | Degree of Acceptance |
|---|---|---|
| A manager appointed for each web project (D4) | 2.91 | Neutral |
| Training program required for all newly-appointed web managers (D16) | 1.65 | Strongly Disagree |

Cluster5 Management Practices: this part is related with project management and training programs. Results (Table 7) illustrates that the appointed manager for each project have not been management according to best practices. That is, they are not sent to training programs to familiarize them with in-house web management project procedures.

Table 8: Process Practices

| Process Practices | Mean Value | Degree of Acceptance |
|---|---|---|
| Development process coped with time pressure (D1) | 3.80 | Agree |
| Development process clarified the roles and responsibilities of each team member (D2) | 3.57 | Agree |

Cluster6 Process Practices: Respondents were asked if the development process that they follow is able to cope with time pressure and whether the development process states clearly the roles and responsibilities of each team members. Results show that both practices have high degree of acceptance (agree). Table 8 shows the results.

Table 9: Testing Practices

| Testing practices | Mean Value | Degree of Acceptance |
|---|---|---|
| Testing process carried out according to requirement specifications to ensure each component was tested (D8) | 2.30 | Disagree |
| Development team performed the testing process (D9) | 3.57 | Agree |

Cluster7 Testing Practices: Here, the testing process on web application components and person responsible for performing the testing were investigated. Table 9 reveals that most developers did not test all components of web applications during the development process. It was also found that most developers perform the testing themselves. This means that developers do not perform the testing process as according to the testing best practices.

## 4. Conclusion

This paper describes current web application development and measurement practices in small software firms in Jordon. The findings show that many developers inside the targeted companies did not follow measurement best practices during the development process. Using the hierarchal clustering technique, seven clusters were identified: requirements practices, quality practices, measurement practice, design practices, management practices, development process practices and testing practices. Results show that from all seventeen practices, twelve practices have not been applied by developers, two practices have neutral acceptance, and three practices have been applied by the developers.

The practices which have not been applied during the development process are related with requirements, quality, measurement, design and testing. These practices are very important to deliver high quality web applications with minimum cost, efforts and short development life cycle.

In general, results from this paper reveal that there is a lack of awareness of applying a set of important development practices during the development process. This clarifies the need for a new web application methodology that integrates some suitable measurement elements. The new methodology will ensure a quality web application product.

## References

[1] J. Conallen, "Modeling Web application architectures with UML," Communications of the ACM, vol. 42, pp. 63-70, 1999.
[2] D. Rodriguez, R. Harrison, and M. Satpathy, "A generic model and tool support for assessing and improving Web processes," in Proc. IEEE Symposium, 2002, pp. 141-151.
[3] I. Richardson and C. Gresse von Wangenheim, "Guest Editors' Introduction: Why are Small Software Organizations Different?," Software, IEEE, vol. 24, pp. 18-22, 2007.
[4] M. E. Fayad, M. Laitinen, and R. P. Ward, "Thinking objectively: software engineering in the small," Communications of the ACM, vol. 43, pp. 115-118, 2000.
[5] C. Hofer, "Software development in Austria: results of an empirical study among small and very small enterprises,"in Proc. (EUROMICRO'02), 2002. pp. 361- 366.
[6] C. Y. Laporte, A. Renault, J. Desharnais, N. Habra, M. Abou El Fattah, and J. Bamba, "Initiating software process improvement in small enterprises: Experiment with micro-evaluation framework,". In Proc. SWDC-REK, 2005, pp. 153–163.
[7] J. McCurley, D. Zubrow, and C. Dekkers, "Measures and measurement for secure software development," Carnegie Mellon University Build Security In, 2008.
[8] A. El Sheikh and H. Tarawneh, "A survey of web engineering practice in small Jordanian web development firms," .in Proc. ACM SIGSOFT, 2007, pp. 481-490.
[9] R. KETTELERIJ. "Designing a measurement program for software development projects", master thesis 2006, Faculty of Science, University of Amsterdam, www.science.uva.nl.
[10] D. H. Withers, "Software engineering best practices applied to the modeling process," Proceedings of the Winter Simulation Conference, Orlando, 2000, pp. 432-439 vol. 1.
[11] Cater-Steel, "An evaluation of software development practice and assessment-based process improvement in small software development firms", PhD thesis, Griffith University, December 2004.
[12] M. Azuma and D. Mole, "Software management practice and metrics in the european community and japan: Some results of a survey," Journal of Systems and Software, vol. 26, pp. 5-18, 1994.
[13] J. D. Blackburn, G. D. Scudder, and L. N. Van Wassenhove, "Improving speed and productivity of software development: a global survey of software developers," Software Engineering, IEEE Transactions on, vol. 22, no.12, pp. 875-885, 1996.
[14] Y. Yasrina, "The Use of Information System Development Methodology in Malaysia". Jurnal Antarabangsa (teknologi maklumat), vol. 2, no. 2002, pp.15-34.
[15] A. McDonald and R. Welland, "A survey of web engineering in practice," Department of Computing Science Technical Report R-2001-79, University of Glasgow, Scotland, vol. 1, 2001
[16] F. Baharom, A. Deraman, and A. Hamdan, "A Survey on the Current Practices of Software Development Process in Malaysia," Journal of ICT, vol. 4, pp. 57–76, 2006.
.[17] NOAA Coastal Services Center, "Introduction to Survey Design and Delivery," Charleston: NOAA Coastal Services Center, 2007.
[18] D. Nachmias, C. Nachmias, and C. F. Nachmias, Research methods in the social sciences: 5th Edition, Aenold a member of the Hodder Headline Group London, 1996.
[19] W. Huang, R. Li, C. Maple, H. Yang, D. Foskett, and V. Cleaver, "Web Application Development Lifecycle for Small Medium-Sized Enterprises (SMEs)," in Proc Of The Eighth International Conference on Quality Software ,2008, pp. 247-252.
[20] J. Pallant, SPSS Survival Manual: A Step by Step Guide to Data Analysis Using SPSS for Windows Version 15: Open University Press, 2007.
[21] Tabachnick, B., & Fidell, L. "Using multivariate analysis." 5th Edition Allyn & Bacon; Needham Heights MA. 2007.
[22] C. Chatfield and A. J. Collins, Introduction to multivariate analysis vol. 166: Chapman & Hall/CRC,London, 1990.
[23] Johnson, R.A. and Wichern, D.W. Applied Multivariate Statistical Analysis. Prentice-Hall of India Private Limited. 4th edition, 1996.
[24] F. Ahmad, F. Baharom and M. Husni, "Investigating the Awareness of Applying the Important Web Application Development and Measurement Practices in Small Software Firms," vol. 3, no.6, pp. 147-158, 2011

**Faudziah Ahmad** She is a senior lecturer at the School of Computing, Universiti Utara Malaysia. She received her Ph.D from Universiti Kebangsaan Malaysia in 2006. She is currently a member of the JICT editorial board. Her research interest includes the data mining, intelligent system, business intelligence and intelligent database. She has been involved in several researches under the Malaysian grants since 1995. Currently, she is supervising several international Ph.D students and local.

**Fauziah Baharom** received her Bachelor in Computer Science from Universiti Teknologi Malaysia in 1992, Master of Science in Software Systems Technology from University of Sheffield, United Kingdom in 1995, and Ph.D. in Computer Science from Universiti Kebangsaan Malaysia in 2008. She works as a Senior Lecturer at the School of Computing, Universiti Utara Malaysia and is the Director for Research Institute of Computing and Technology. She is an active researcher and has lead several research grants. Her research interest includes Software Process Quality, Improvement and Evaluation; Secure Software Development; Software Certification and e-Government survivability.

**Moath Husni** is a Ph.D. candidate in Universiti Utara Malaysia (UUM).